\documentclass[a4paper,11pt]{article}
\usepackage{pos}
\usepackage{float}
\usepackage{subfigure}  
\usepackage{siunitx}
\usepackage{booktabs}
\usepackage{mathtools}
\usepackage{physics}
\usepackage{hyperref}   
\usepackage{bbm}
\usepackage[noabbrev, nameinlink]{cleveref}
\usepackage{layouts}
\usepackage{pgfplots}
\usepgfplotslibrary[groupplots,dateplot]
\usetikzlibrary{patterns,shapes.arrows}
\pgfplotsset{compat=newest}
\title{Metadynamics Surfing on Topology Barriers \\ in the Schwinger Model}
\ShortTitle{Metadynamics Surfing on Topology Barriers in the Schwinger Model}

\author{Timo Eichhorn}
\author{Christian Hoelbling}
\author*{Philip Rouenhoff}
\author{Lukas Varnhorst}

\affiliation{Department of Physics, University of Wuppertal, Gaußstraße 20, D-42119 Wuppertal, Germany}

\emailAdd{philip.rouenhoff@uni-wuppertal.de}

\abstract{Topological freezing is a well known problem in lattice simulations: with shrinking lattice spacing a transition between topological sectors becomes increasingly improbable, leading to a problematic increase of the autocorrelation time regarding several observables. We present our investigation of metadynamics as a solution for topological freezing in the Schwinger model. Specifically, we take a closer look at the collective variable and its scaling behaviour, visualize the effects of topological freezing and how metadynamics helps in that respect and explore alternatives for a more efficient building process. Possible implications for and differences to four-dimensional SU(3) theory are briefly discussed.}

\FullConference{
 The 39th International Symposium on Lattice Field Theory, LATTICE2022
  8th-13th August, 2022
  Bonn Germany
}

\begin{document}
\maketitle
\newcommand\thefont{\expandafter\string\the\font}

\section{Introduction}
In order to generate configurations in an efficient manner, one has to rely on the method of importance sampling, the reason being that the probability of generating relevant configurations via simple sampling decreases drastically for increasing dimensions of the phase space. Importance sampling is usually implemented by means of Markov chain Monte Carlo algorithms, which per se involve generating configurations that are correlated with each other. For several interesting field theories (such as 2-dim. U(1) or 4-dim. SU(3) gauge field theories) one notices a dramatic increase of the autocorrelation time when approaching the continuum limit. This phenomenon, which is called topological freezing, ultimately thwarts correct measurements of observables and still poses an active topic of research \cite{Alles:1996vn, deForcrand:1997fm, DelDebbio:2004xh, Schaefer:2009xx, Durr:2012te}.

\begin{figure}[H]
    \centering
    \includegraphics[width=\textwidth]{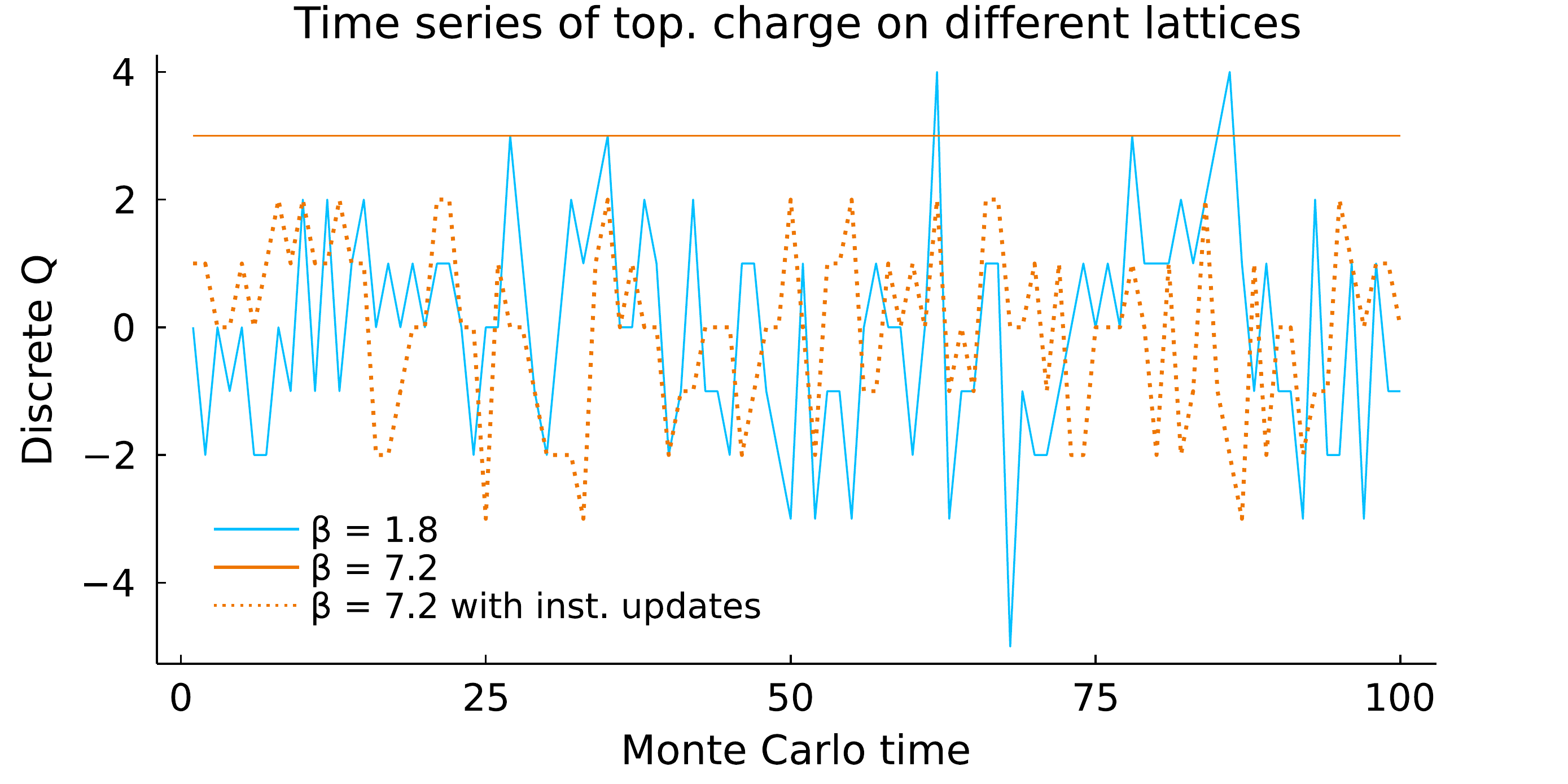}
    \caption{Time series of the discrete topological charge on different square lattices. The configurations were produced using the Metropolis algorithm on a line of constant physics (LCP) given by $N_x N_t / \beta = 80$. On the finer lattice with $\beta = 7.2$ the system visibly got stuck in a topological sector. Including instanton-updates in between Metropolis sweeps caused the system to tunnel to different sectors again, providing a remedy. }
    \label{fig:topological_freezing}
\end{figure}

One observable which is particularly prone to topological freezing is the topological charge $Q$, as can be seen in Fig.\,\ref{fig:topological_freezing}. It can be defined in both 2-dim. U(1) and 4-dim. SU(3) field theory. We will focus on the former case, where we can luxuriously define $Q$ in a manner which only yields integer values:
\begin{equation}
    Q = \frac{1}{2\pi} \sum_{\Vec{n}\in\Lambda} \mathfrak{Im}\Big[\log(P_{tx}(\Vec{n}))\Big],
\end{equation}
where $P_{tx}(\Vec{n})$ is the plaquette at the lattice site $\Vec{n} = (t,x)^\text{T}$. Since small changes of a configuration do not always change $Q$, there are regions in phase space where $Q$ is constant, called topological sectors. The troublesome increase in autocorrelation time is caused by action barriers in between these sectors that grow for decreasing lattice spacing $a$, ultimately trapping the Markov chain inside. The method of Metadynamics helps visualize and circumvent these action barriers, as can be seen in the next section. \vspace{15pt}


\section{Metadynamics}

For 2-dim. U(1) theory there already exist multiple methods which counteract topological freezing, one example being instanton-updates \cite{Smit:1986fn, Belavin:1975fg, Durr:2012te} which involve multiplying a configuration link by link with a $\pm1$-instanton (plus or minus with equal probability) followed by a conventional accept-reject step. Here a $Q$-instanton is a configuration given by
\begin{equation}\label{eq:insta}
\begin{aligned}
    U_{t}^{I}(Q; t, x) &= \exp\left( -2 \pi i x \frac{ Q}{N_{t} N_{x}} \right),
    \\U_{x}^{I}(Q; t, x) &= \exp\left( 2 \pi i t \frac{Q}{N_{t}} \delta_{x, N_{x}} \right).
\end{aligned}
\end{equation}
Consequently, every plaquette of a $Q$-instanton configuration has the same value, such that its topological charge is $Q$. Thus, an instanton-update proposes a configuration whose topological charge differs by $\Delta Q = \pm 1$, effectively tunneling through the action barriers. In 2-dim. U(1) theory this update is very effective when used in combination with ergodic algorithms such as the Metropolis algorithm used here, as can be seen in Fig.\,\ref{fig:topological_freezing}. However, in 4-dim. SU(3) there are multiple problems, which also holds for other topology changing algorithms \cite{Eichhorn:2021ccz, Eichhorn:2022wxn}.\\

Metadynamics is a topology changing algorithm that also seems promising for SU(3) \cite{Eichhorn:2021ccz, Eichhorn:2022wxn}. It involves building a bias potential $V$ (also called metapotential) that depends on so called collective variables (CV) and is added onto the gauge action. The idea is to add small and local repulsive potentials at the points in phase space (parameterized via the CVs) that the system has already visited, thus discouraging the system from revisiting the same places again and eventually filling up local action minima. 
Observables can be measured via reweighting with factors $e^{V_i}$, see Sec.\,\ref{sec:ess}.\\

\begin{figure}[h]
    \centering
    \subfigure[$32\times 32$-lattice with $\beta = 12.8$]{
        \includegraphics[width=0.34\paperwidth]{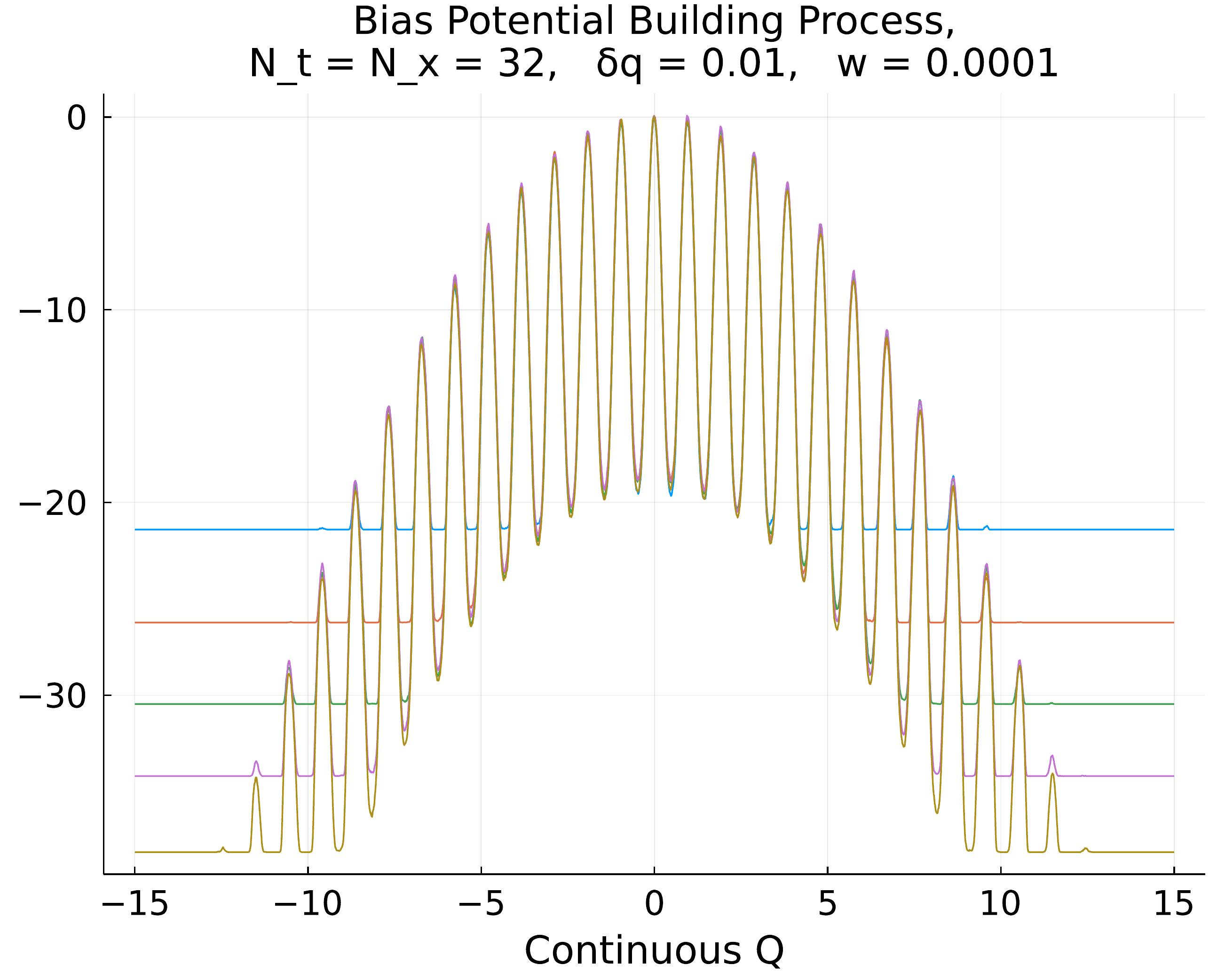}
        \label{fig:bias_potential_b}
    }
    \subfigure[Measurement of $\langle Q^2\rangle$ for different $\beta$]{
        \includegraphics[width=0.34\paperwidth]{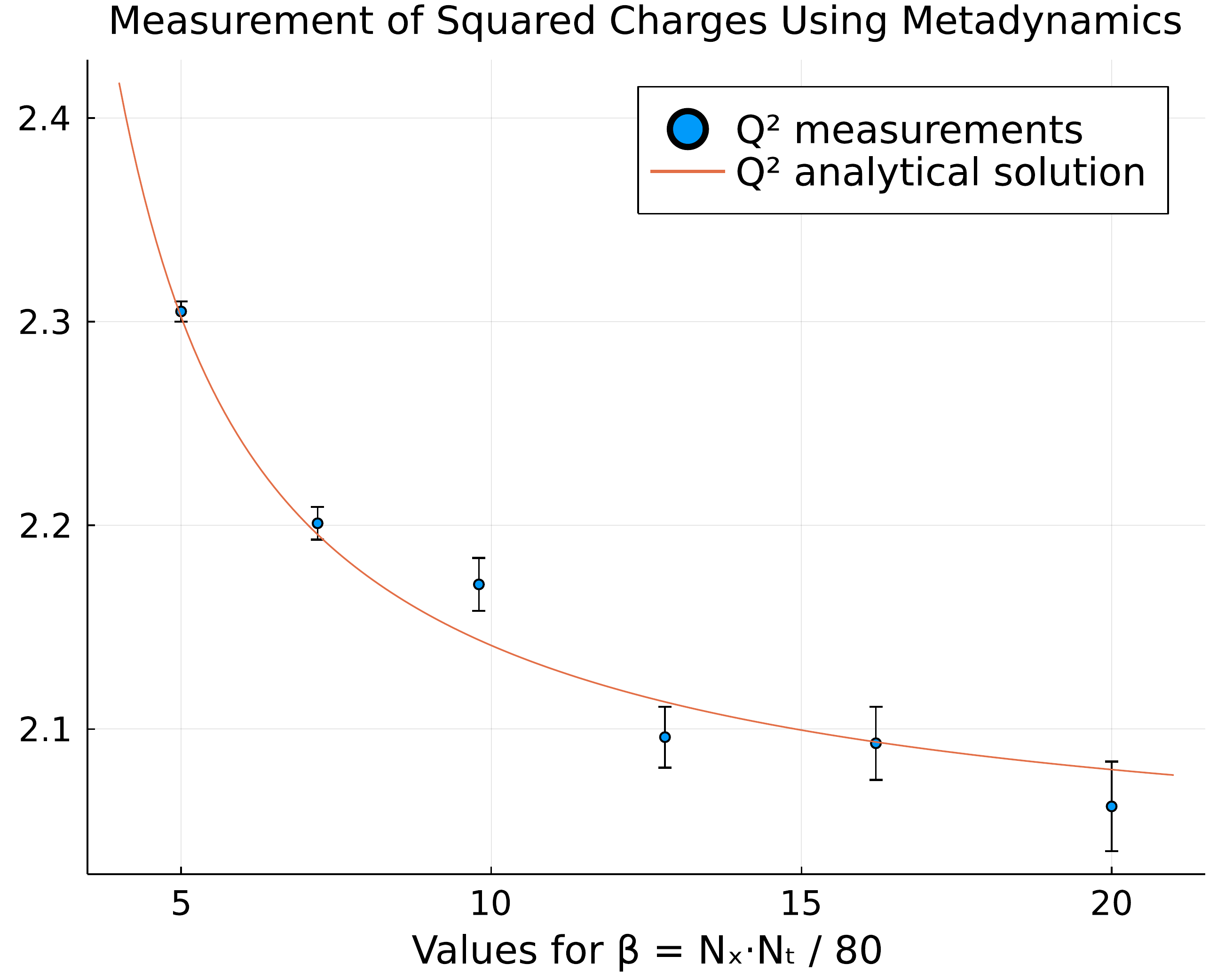}
        \label{fig:bias_potential_c}
    }
    \caption[Examples of the building process of metapotentials]{ \subref{fig:bias_potential_b} Shows snapshots of the metapotential taken at different times during the building process. For comparison the maximum value was subtracted at each time. In \subref{fig:bias_potential_c} measurements of $\langle Q^2 \rangle$ using Metadynamics are compared with the analytical prediction \cite{Elser:2001pe}. The $p$-value is $p = 97.8\%$ and $\chi^2/\text{dof} = 7.02/6 = 1.17$. }
    \label{fig:bias_potential}
\end{figure}

To be more specific we proceed analogously to Laio et. al \cite{Laio:2015era}. We use one CV to characterize the phase space, which we call the continuous topological charge $Q_{\text{cont}}$ (also called meta charge and denoted by $Q_\text{meta}$), as it is an approximation of the discrete charge and not integer-valued anymore:
\begin{equation}\label{eq:cont_charge}
    Q_{\text{cont}} = \frac{1}{2\pi} \sum_{\Vec{n}\in\Lambda} \mathfrak{Im} \Big[P_{tx}(\Vec{n})\Big].
\end{equation}
$Q$ itself, telling us in which sector the system is currently located, is already a good means of characterizing the phase space. For Metadynamics, however, it is necessary to have a higher resolution, which is why $Q_{\text{cont}}$ is the CV of choice here. In order to build up the bias potential, one starts a run using the Metropolis algorithm, 
measures the CV at each point $t$ in Monte Carlo time and adds a small strictly positive potential $g(Q_{\text{cont}}(x(t)))$ onto $V$. Thus, $V$ is built up according to
\begin{equation}\label{eq:potential}
    V(Q_{\text{cont}}(x), t) = \sum_{t' < t} g\Big(Q_{\text{cont}}(x) - Q_{\text{cont}}\big(x(t')\big)\Big),
\end{equation}
where $x$ is a point in phase space. The potential is stored on a $Q_\text{cont}$-grid of resolution $\delta Q$. 
It is important that $g$ vanishes rapidly for large absolute values of its argument; we used triangles of height $w$, which is a little cheaper than e.g. $g(Q) = w\cdot\exp(\frac{-Q^2}{2\delta Q})$, which has also been used before \cite{Laio:2015era}. An example of a bias potential as well as a demonstration of principle can be seen in Fig.\,\ref{fig:bias_potential}. The values used here are $w = 10^{-4}$ and $\delta Q = 10^{-2}$.

\section{Renormalization Constant $Z$}

When plotting the bias potential as in Fig.\,\ref{fig:bias_potential_b}, the extrema apparently do not align with the integer values on the $Q_\text{cont}$-axis. This is to be expected, since the local action minima lie at $Q$-instanton configurations, and for their continuous and discrete charges one can quickly see by use of Eqs.\,\eqref{eq:insta} and \eqref{eq:cont_charge} that for fixed instanton charge $Q_\text{insta}$ holds $|Q_\text{cont, insta}| \leq |Q_\text{insta}|$. \footnote{In fact, one can swiftly calculate that $Q_\text{cont, insta}$ has a sinusoidal behaviour with a period of $(N_x N_t)^{-1}$.}

\begin{figure}[h]
    \centering
    \subfigure[2D histograms of $(Q, Q_{\text{cont}})$-pairs]{
        \includegraphics[width=0.34\paperwidth]{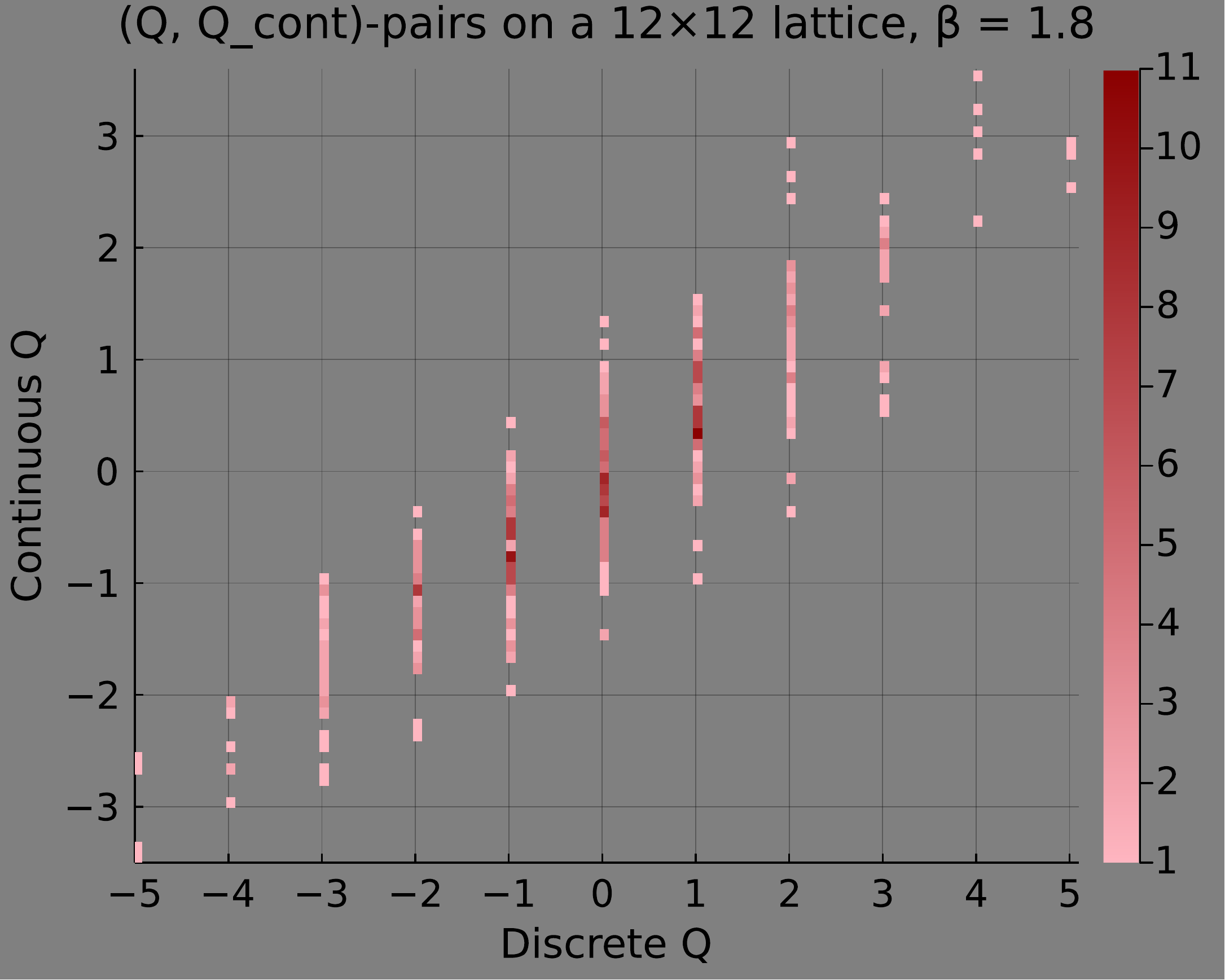}
        \label{fig:q_dis_vs_q_cont_a}
    }
    \subfigure[Linear fit through $(Q, Q_{\text{cont}})$-pairs]{
        \includegraphics[width=0.34\paperwidth]{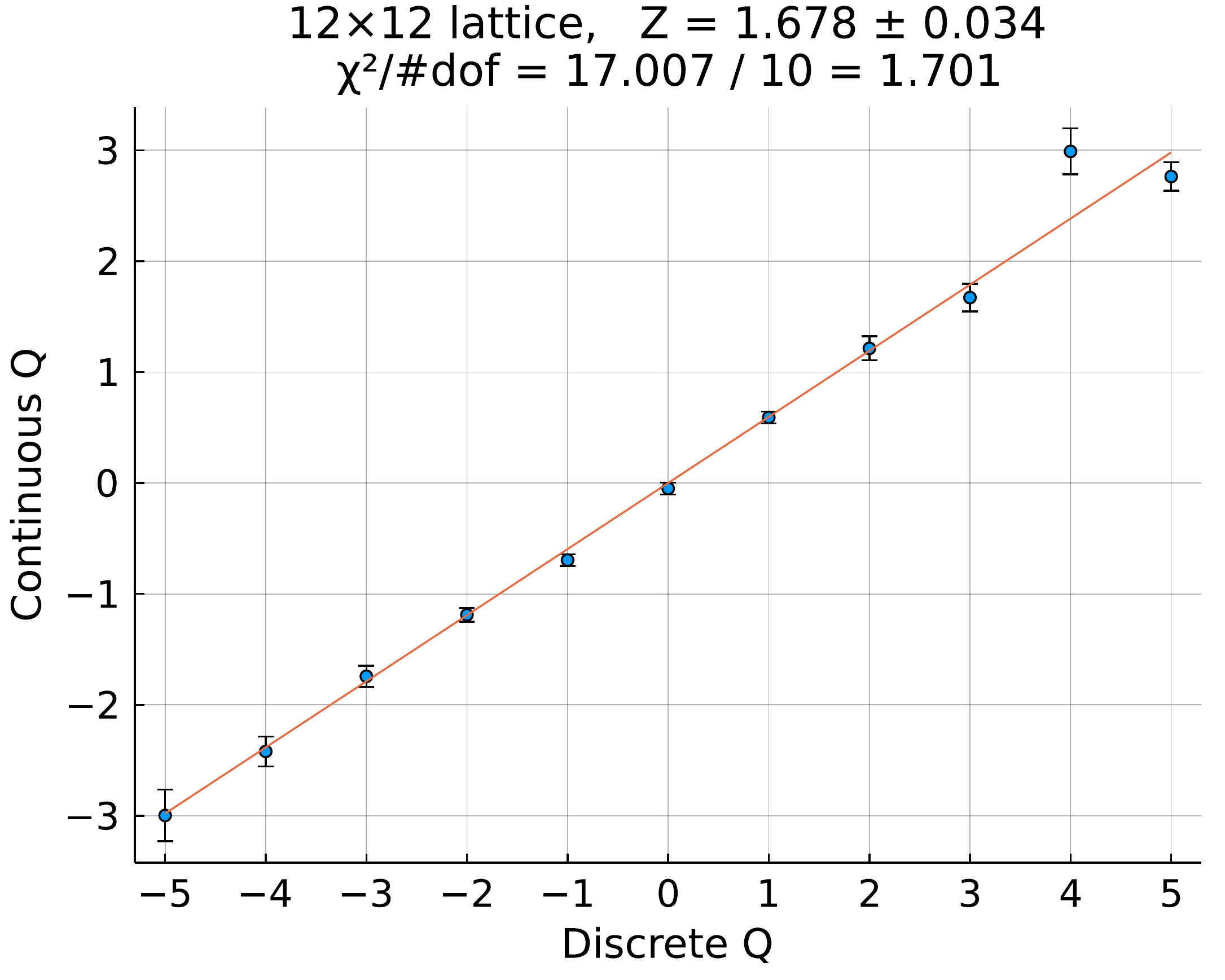}
        \label{fig:q_dis_vs_q_cont_b}
    }
    \caption[Comparison of $Q_{\text{dis}}$ and $Q_{\text{cont}}$ on different lattices]{Measurements on a $(12\times 12)$-lattice with $\beta = 1.8$. This is the coarsest lattice used, for finer lattices the widths of the distributions in \subref{fig:q_dis_vs_q_cont_a} quickly decrease while the slopes in \subref{fig:q_dis_vs_q_cont_b} approach one. }
    \label{fig:q_dis_vs_q_cont}
\end{figure}

We measured pairs of $(Q,Q_\text{cont})$ of configurations generated with the Metropolis algorithm infused with instanton-updates on various lattice sizes and plotted them in 2D histograms as in Fig.\,\ref{fig:q_dis_vs_q_cont_a}. The means of the $Q_\text{cont}$-distributions corresponding to one $Q$-value each could be fitted by a linear function of $Q$, see Fig.\,\ref{fig:q_dis_vs_q_cont_b}.

\begin{figure}[h]
    \centering
    \subfigure[Without multiplying with $Z$]{
        \includegraphics[width=0.34\paperwidth]{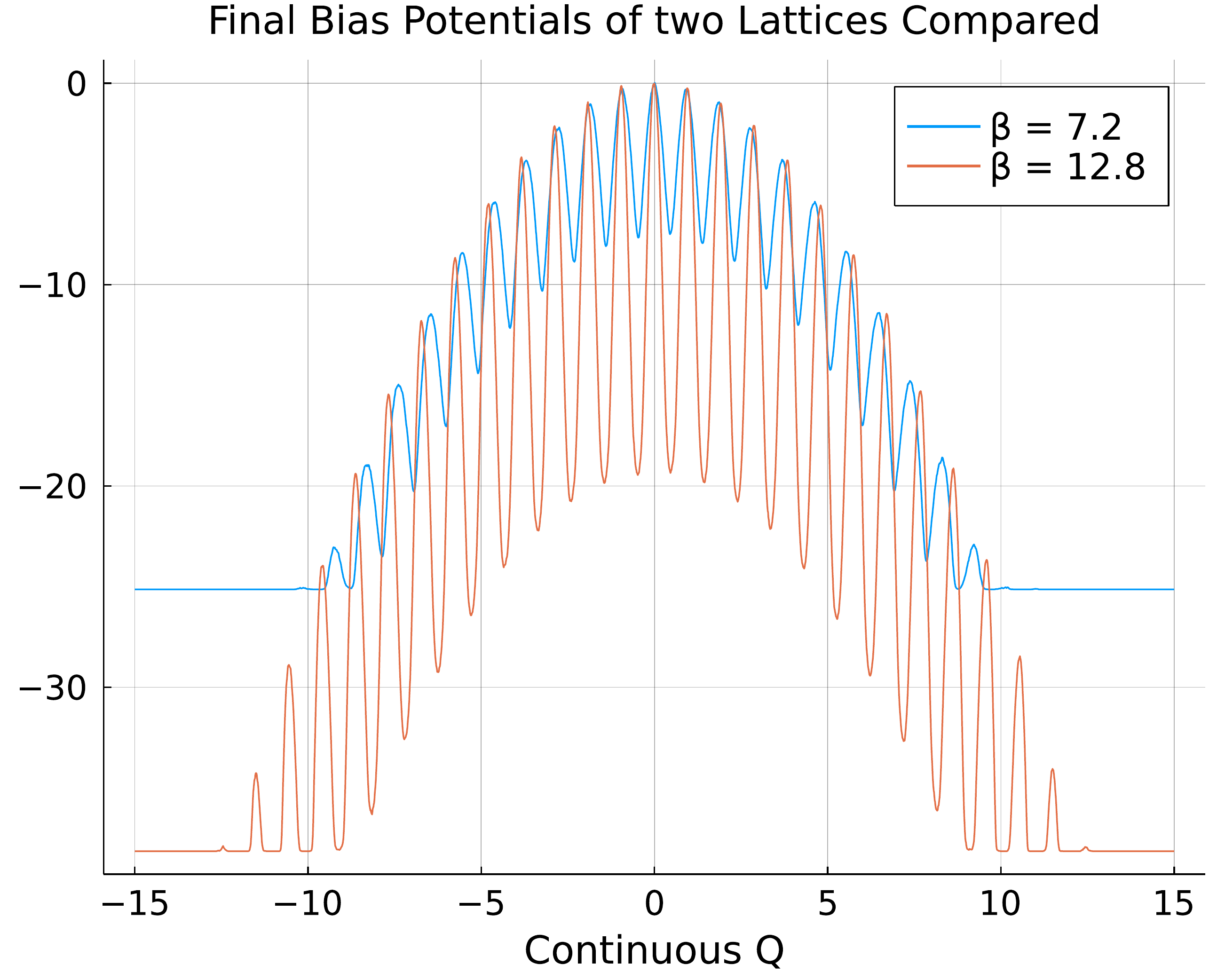}
        \label{fig:biases_a}
    }
    \subfigure[With multiplying with $Z$]{
        \includegraphics[width=0.34\paperwidth]{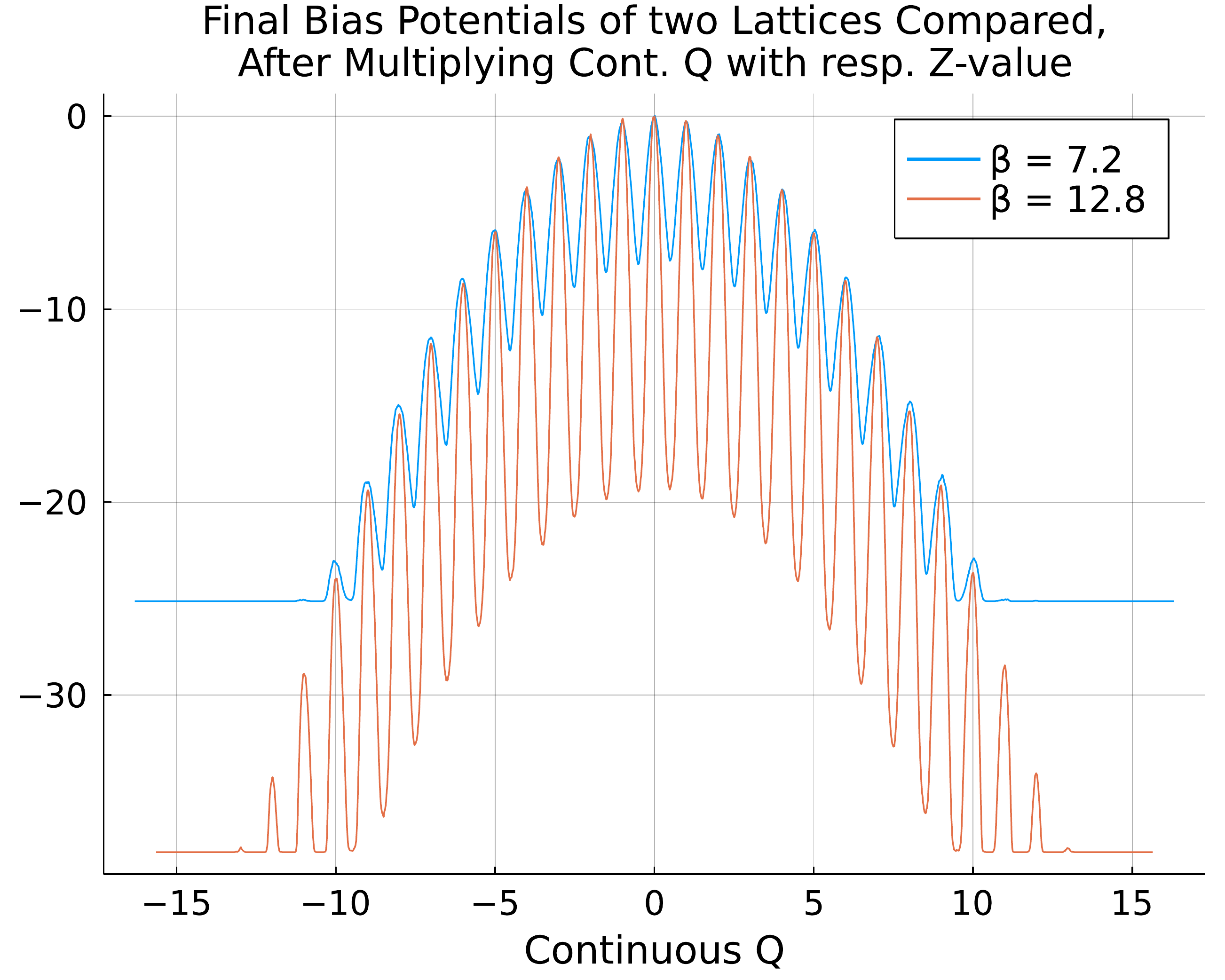}
        \label{fig:biases_b}
    }
    \caption[Comparison of two final bias potentials and demonstration of the use of $Z$]{ Due to the scaling of the continuous charge one cannot directly compare metapotentials of different $\beta$'s of the same LCP: one has to multiply the $Q_{\text{cont}}$-values with their respective $Z$-factors.}
    \label{fig:biases}
\end{figure}

As a consequence of the definition of $Q_\text{cont}$, the slope in Fig.\,\ref{fig:q_dis_vs_q_cont_b} is less than one. We call its inverse the renormalization constant $Z$, since multiplying $Q_\text{cont}$ with $Z$ leads to the mean values of the $Q_\text{cont}$-distributions aligning with their respective integer values. Hence, considering $Z\cdot Q_\text{cont}$ instead of $Q_\text{cont}$ leads to the extrema of the bias potential aligning with (half-) integer values as can be seen in Fig.\,\ref{fig:biases}.

The fitting function we found to describe $Z$ best as a function of the lattice spacing $a$ is
\begin{equation}\label{eq:fit}
    Z_{\,\text{fit}}(a) = (1.001 \pm 0.001) + (33.50 \pm 2.78)\, a^2 + (9431 \pm 866)\, a^4,
\end{equation}
although, apart from polynomial functions, also exponential and Pad\'e ansatzes have been looked into. Note that the limit $\lim_{a\rightarrow 0} Z(a) = 1$ is an important check for consistency as $Q$ and $Q_\text{cont}$ both have the same continuum limit. The dependency of $Z$ on only even powers of $a$ is to be expected since the presence of $a$ raised to uneven powers would break the symmetry of the topological charge distribution.


\section{Effective Sample Size}\label{sec:ess}

Modifying the action with the bias potential results in sampling configurations according to a different probability distribution than that of the underlying theory. Consequently, as mentioned before, one has to make use of reweighting when measuring observables. In our case, the weights are obtained using the bias potential entries of the $i$-th ensemble member via $w_i = \exp(V(Q_{\text{cont},i}))$, leading to
\begin{equation}
    \langle O \rangle = \frac{\sum_i O_i \exp(V(Q_{\text{cont},i}))}{\sum_i \exp(V(Q_{\text{cont},i}))}.
\end{equation}
This results in a different effective sampling size $n_\text{eff}$, which can be calculated via
\begin{equation}\label{eq:ess}
    n_\text{eff} = \frac{\left(\sum_i w_i \right)^2 }{\sum_i w_i^2}.
\end{equation}
This has been done for the bias potentials on various lattice sizes, the results can be found in Tab.\,\ref{tab:ess}.

\begin{table}[H]
    \centering
    \begin{tabular}{|c|c|c|c|c|} \hline
        $N$ & $\beta$   & $n_\text{eff}$    & ratio         & $\tau_{\textrm{int}}$ \\\hline\hline
        20  & 5.0       & 1955671           & 0.19559       & 7357  \\\hline
        24  & 7.2       & 941015            & 0.09411       & 13881 \\\hline
        28  & 9.8       & 643229            & 0.06433       & 12832 \\\hline
        32  & 12.8      & 614815            & 0.06149       & 12605 \\\hline
        36  & 16.2      & 307691            & 0.03077       & 17518 \\\hline
        40  & 20.0      & 182351            & 0.01824       & 33192 \\\hline
    \end{tabular}
    \caption{Effective sample sizes as per Eq.\,\eqref{eq:ess}. $10^7$ configurations were generated on lattices of different $N = N_t = N_x$ on the same LCP. The integrated autocorrelation time was determined using $\langle Q^2 \rangle$. }
    \label{tab:ess}
\end{table}

To avoid sampling regions of unnecessarily high $Q_\text{cont}$, a penalty potential was used such that the generation of configurations with $|Q_\text{cont}| \geq 7.0$ was heavily suppressed. Yet the measured $n_\text{eff}$ turn out to be comparatively small, highlighting one shortcoming of Metadynamics. To that end one can enhance the procedure by adapting the height $w$ of the small local potentials $g(Q_\text{cont})$ dynamically, i.e. $w \rightarrow w(t)$, such that $w(t)$ decreases over Monte Carlo time. This approach is referred to as well-tempered Metadynamics \cite{Barducci_2008} and has been shown to provide relief.

\section{Fitting Attempts}

\begin{figure}[H]
    \centering
    \subfigure[$20\times 20$-lattice with $\beta = 5.0$]{
        \includegraphics[width=0.34\paperwidth]{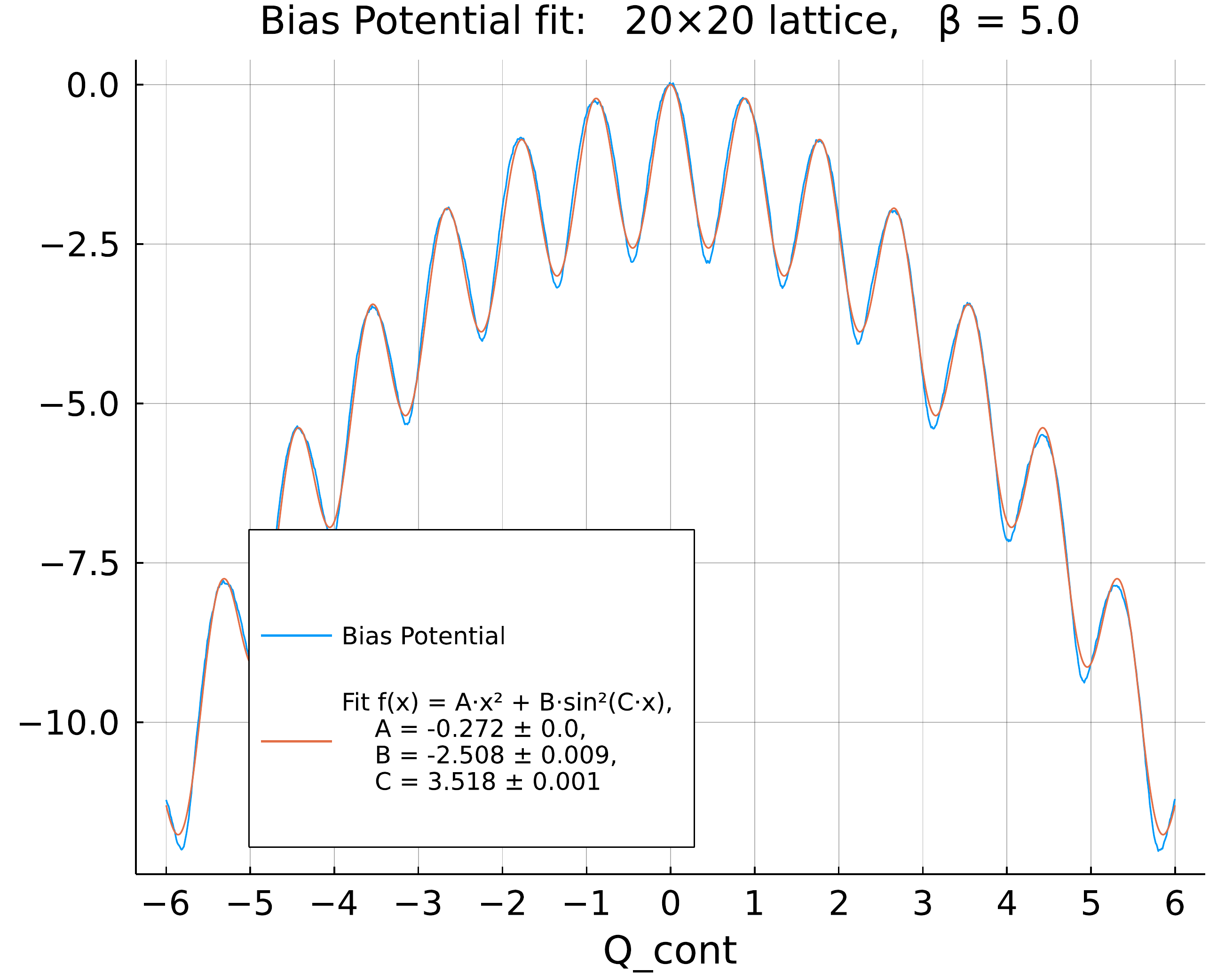}
        \label{fig:first_fit_a}
    }
    \subfigure[$40\times 40$-lattice with $\beta = 20.0$]{
        \includegraphics[width=0.34\paperwidth]{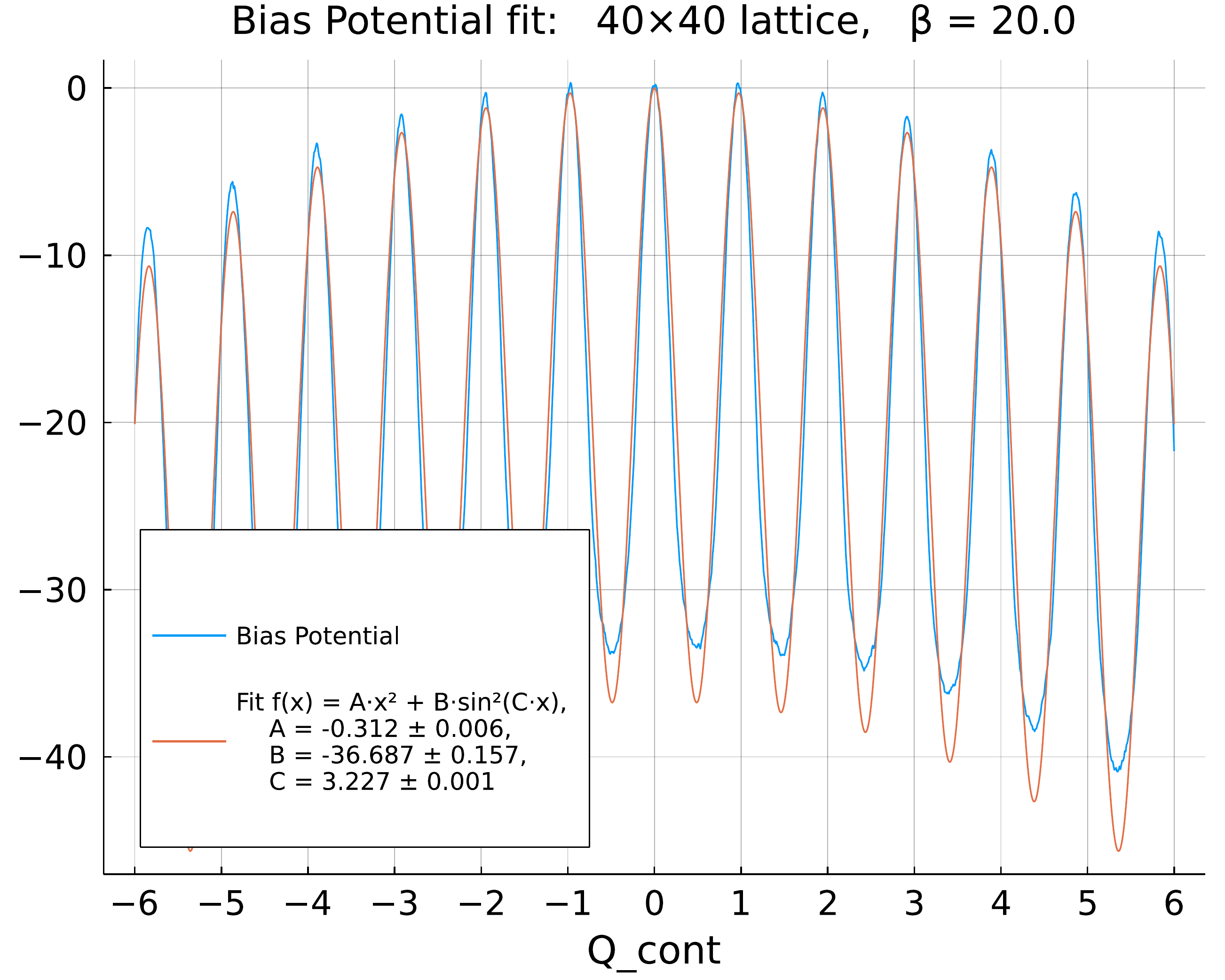}
        \label{fig:first_fit_b}
    }
    \caption[]{Using the naive fitting function $F(Q_\text{cont}) = A Q_\text{cont}^2 + B \sin^2(C Q_\text{cont})$\, to describe the bias potential does not yield satisfactory results for any lattice size. }
    \label{fig:first_fit}
\end{figure}

One drawback of Metadynamics is the circumstance that a bias potential has to be built up 
for every lattice size separately, leading to an increase of computational costs. Knowing the bias potential in advance would thus impose a significant improvement. Even an estimate can be of advantage, since small corrections can be implemented via a shortened building run. We are currently performing fits of already built up bias potentials in hopes of finding a pattern of the fit parameters for different lattice sizes on the same LCP. 

\begin{figure}[H]
    \centering
    \subfigure[]{
        \includegraphics[width=0.34\paperwidth]{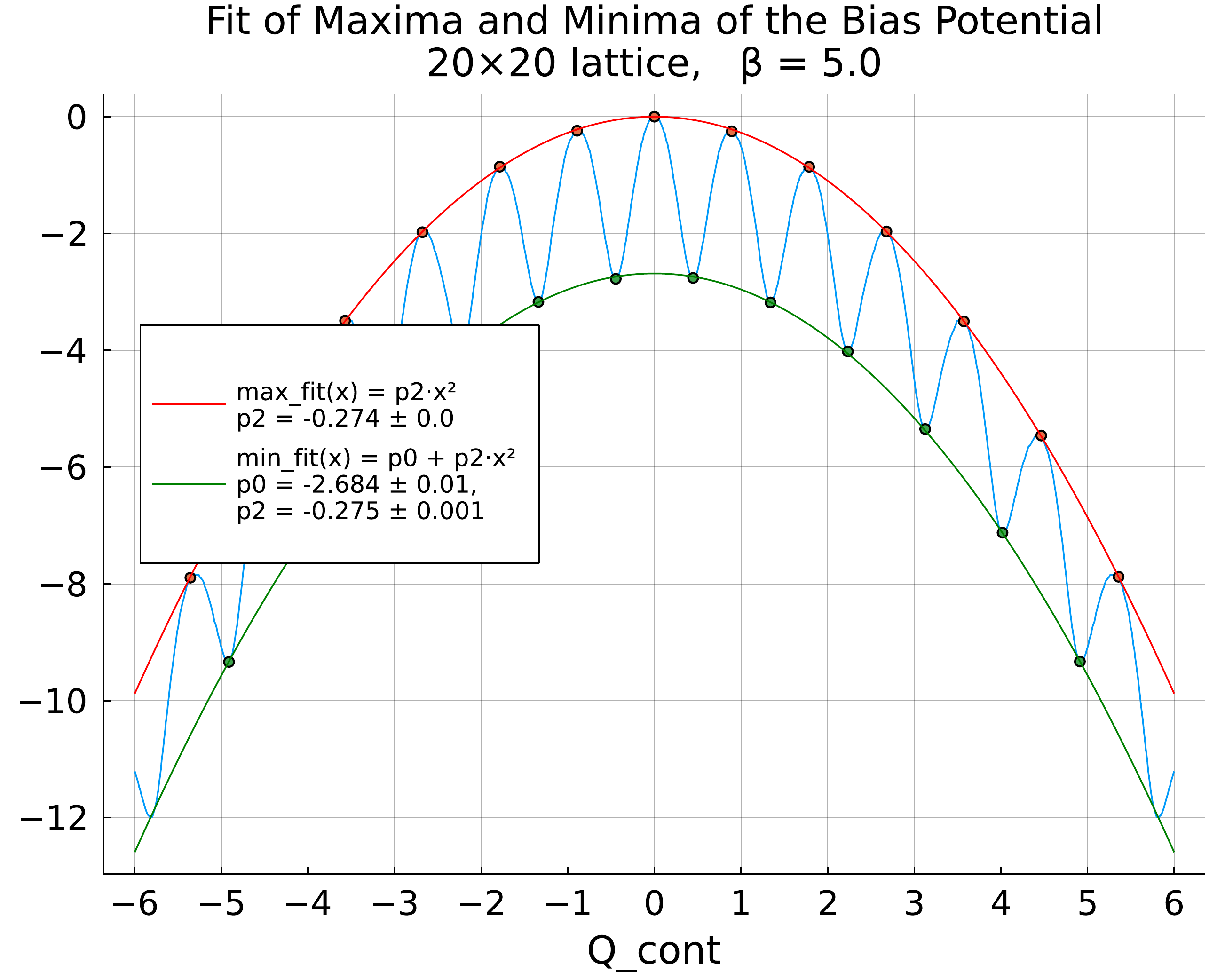}
        \label{fig:fourier_fit_a}
    }
    \subfigure[]{
        \includegraphics[width=0.34\paperwidth]{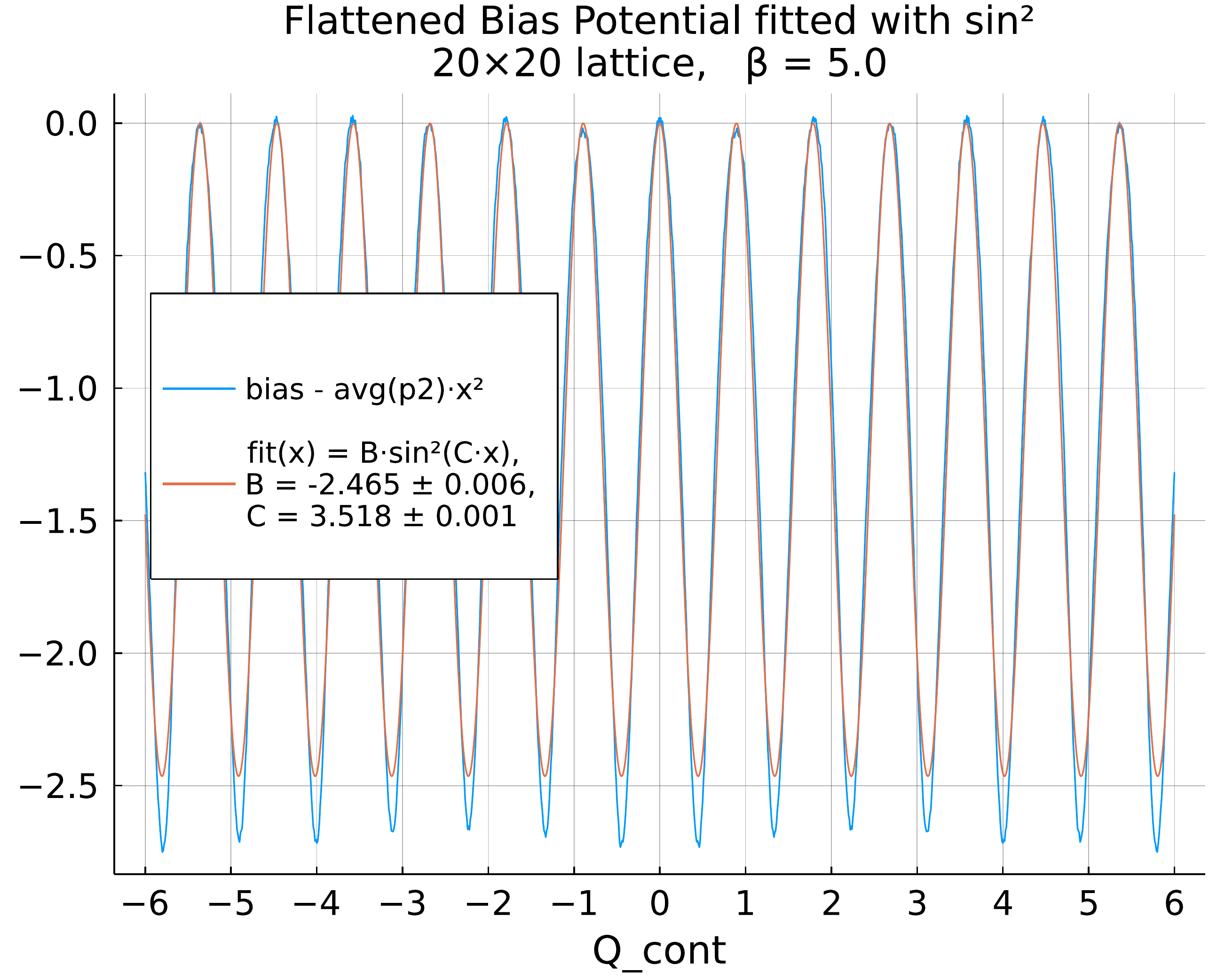}
        \label{fig:fourier_fit_b}
    }
    \subfigure[]{
        \includegraphics[width=0.34\paperwidth]{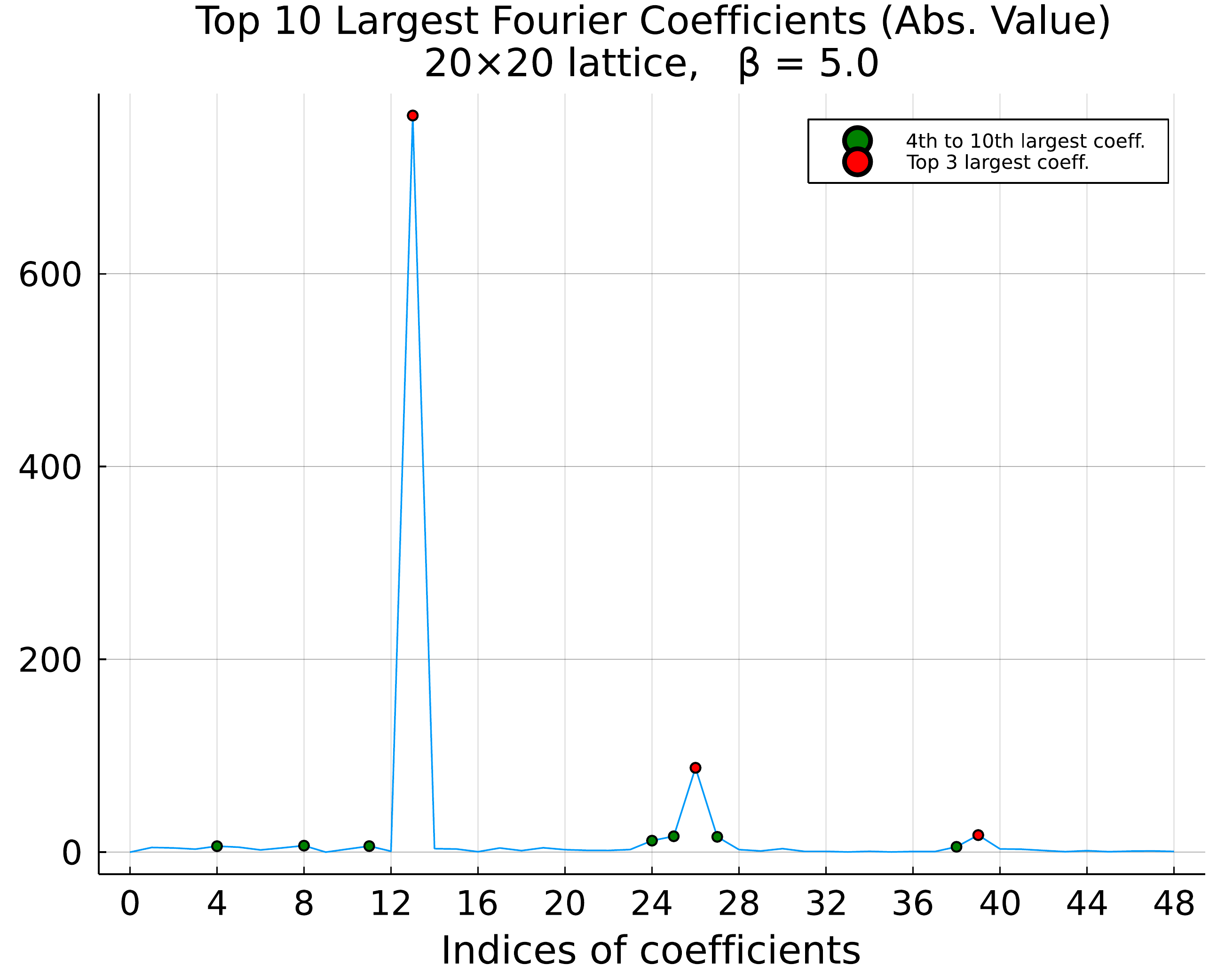}
        \label{fig:fourier_fit_c}
    }
    \subfigure[]{
        \includegraphics[width=0.34\paperwidth]{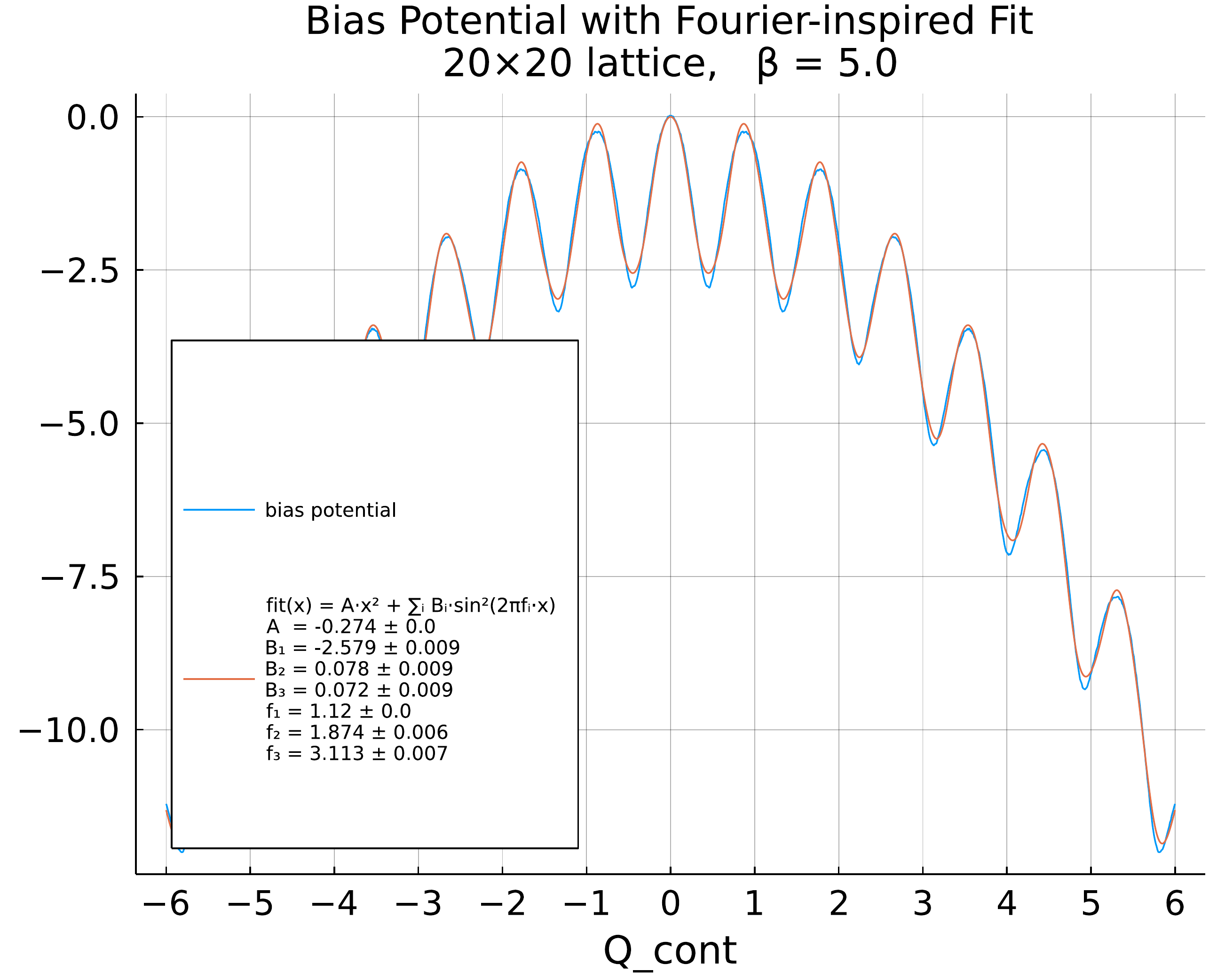}
        \label{fig:fourier_fit_d}
    }
    \caption[]{Illustration of the fitting process using Fourier transforms. First, parabolas $p_2 x^2$ and $p_0 + p_2 x^2$ were fitted to the maxima and minima, respectively, as seen in \subref{fig:fourier_fit_a}. Since their $p_2$-parameters are compatible with each other in almost all of the cases, the parabola avg($p_2) x^2$ was subtracted, which yields the signal seen in \subref{fig:fourier_fit_b}. There one can also see that a single $\sin^2$-term still does not suffice. The most relevant frequencies can be seen in \subref{fig:fourier_fit_c}, which were the proposals for the fit in \subref{fig:fourier_fit_d}. }
    \label{fig:fourier_fit}
\end{figure}

While Laio et al. \cite{Laio:2015era} showed that a fit of the form $F(Q_\text{cont}) = AQ_\text{cont}^2 + B \sin^2(\pi Q_\text{cont})$\, described the model of their choice well, we find that this function is insufficient in 2-dim. U(1)-theory: firstly, the phase velocity must be modified by the $Z$-factor since here the local action minima lay in vicinity of $Q$-instanton configurations, which are defined using the discrete charge, not the continuous one; secondly, in our case it seems that for coarse lattices the fitting function consistently under- and for finer lattices overestimates the potential barriers. Both of these observations are illustrated in Fig.\,\ref{fig:first_fit}.

\begin{figure}[h]
    \centering
    \subfigure[$f_i$ for various lattices]{
        \includegraphics[width=0.34\paperwidth]{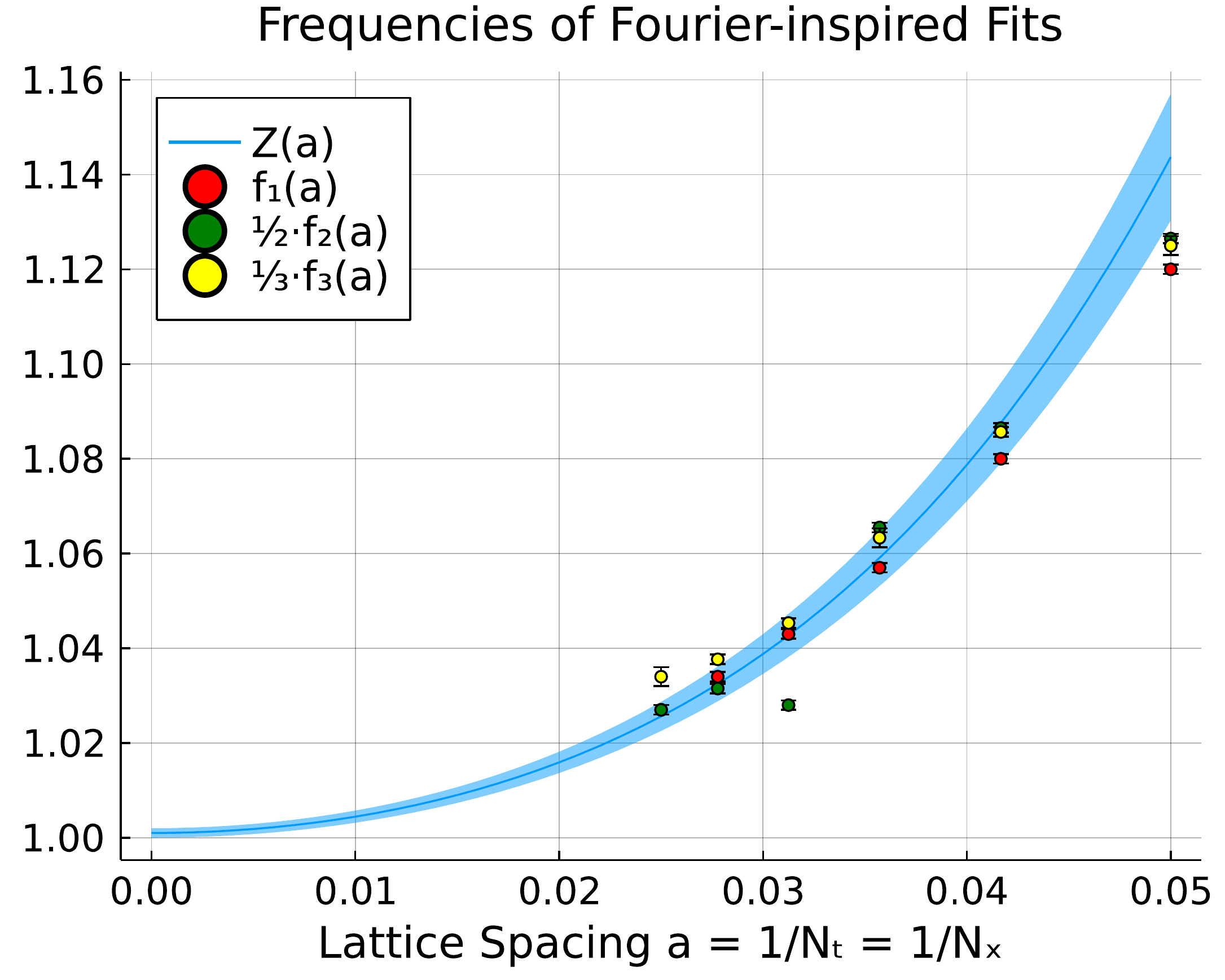}
        \label{fig:fit_results_a}
    }
    \subfigure[Preliminary $B_i$ for various lattices]{
        \includegraphics[width=0.34\paperwidth]{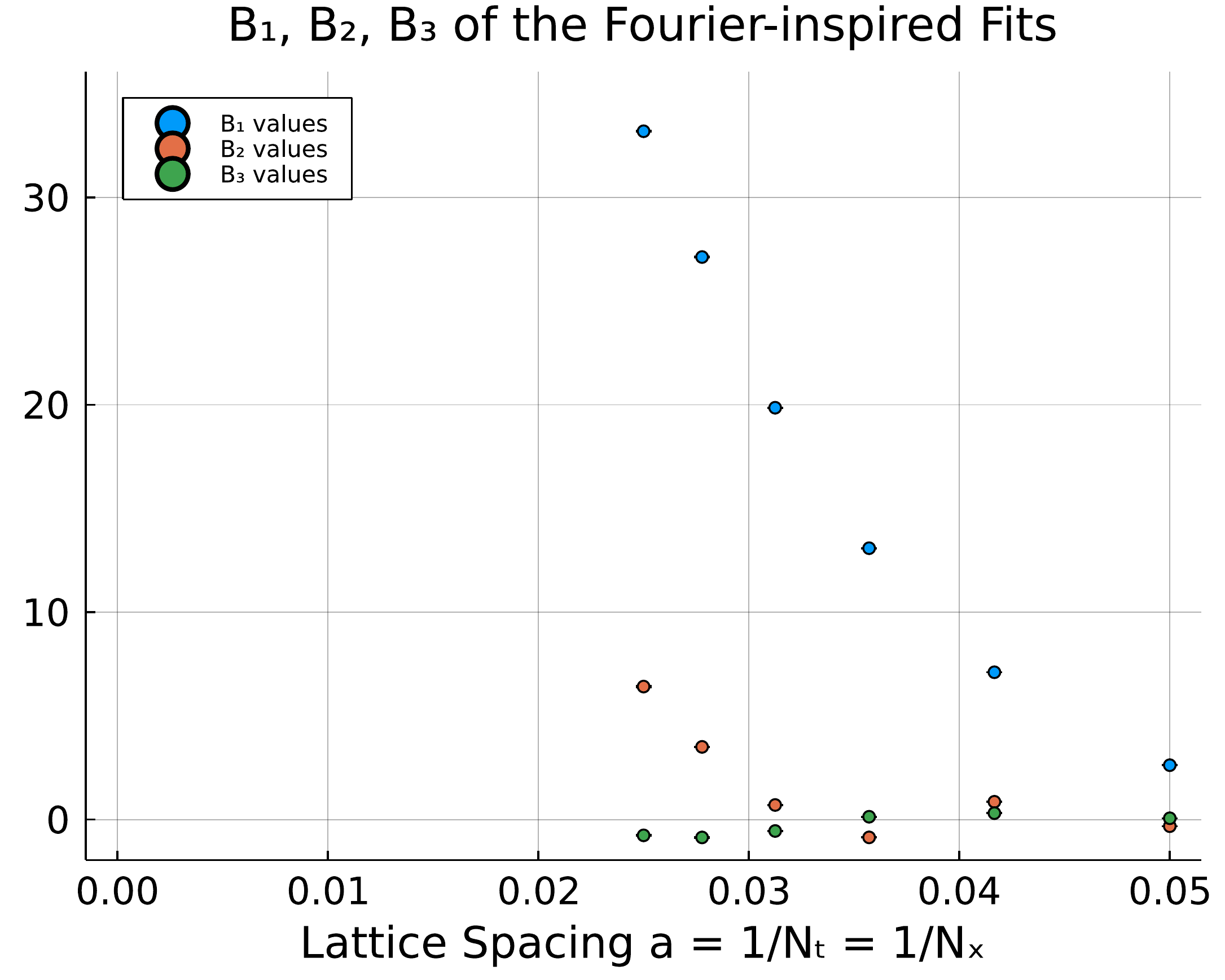}
        \label{fig:fit_results_b}
    }
    \caption[]{In \subref{fig:fit_results_a} the frequencies $f_i$ from the fitting function Eq.\,\eqref{eq:fourier_fit} are plotted over $a$. One can see that for fixed $a$ the frequencies $f_2$ and $f_3$ roughly match the first two harmonics of $f_1$ and additionally, that the frequencies coincide with (multiples of) $Z(a)$ for small $a$. \subref{fig:fit_results_b} shows preliminary results for the amplitudes $B_i$. }
    \label{fig:fit_results}
\end{figure}

Using multiple $\sin^2(\pi f_i Q)$ -terms with frequencies $f_i$ determined via a discrete Fourier transform yields a better result. The fitting process used here is displayed in more detail in Fig.\,\ref{fig:fourier_fit}. To outline, parabolas were fitted to the extrema of the bias potentials in order to subtract them so as to receive a signal more suited for a Fourier transform. The frequency spectrum can be seen in Fig.\,\ref{fig:fourier_fit_c} and turned out to be mostly composed of one ground frequency close to the aforementioned $C$-parameters and its following one to two harmonics. These frequencies were then used as proposals for $f_i$ in fitting functions of the form 
\begin{equation}\label{eq:fourier_fit}
    F'(Q_\text{cont}) = A' Q_\text{cont}^2 + \sum_{i=1}^{2\text{ or }3} B_i \sin^2(\pi f_i Q_\text{cont}),
\end{equation}


\noindent where $A'$, $B_i$ and $f_i$ are fit parameters. We present our results for the case $i\in\{1,2,3\}$ and note that the transition from $i\in\{1,2\}$ to $i\in\{1,2,3\}$ did not produce different results for $B_1$, $f_1$ and $f_2$, but did so for $B_2$. The frequencies determined for different lattice sizes can be seen in Fig.\,\ref{fig:fit_results_a}. There it can be seen that for finer lattices, the frequencies roughly coincide with the prediction delivered by the renormalization constant $Z(a)$.

Fig.\,\ref{fig:fit_results_b} shows the preliminary results for the fitted amplitudes $B_i$. The growth of $B_1$ for decreasing $a$ clearly shows the growth of the action barriers when approaching the continuum, as we already knew. For $B_2$ and $B_3$ one can see that while for coarse lattices both of the amplitudes are comparatively small, for finer lattices $B_2$ seems to grow. One can surmise that for finer lattices $B_3$ will follow this behaviour and become more relevant as well.


\section{Conclusion}

In 2-dim. U(1) gauge theory the approach of Metadynamics seems to be a successful remedy for topological freezing. One drawback of this method, the building process, is attempted to be circumvented by the search of suitable fitting functions for the bias potentials. One function that yields a satisfactory result is
\begin{equation*}
    F'(Q_\text{cont}) = A' Q_\text{cont}^2 + \sum_{i=1}^{2\text{ or }3} B_i \sin^2(\pi f_i Q_\text{cont}),
\end{equation*}
where a Fourier analysis of the bias potential showed that the frequencies $f_1, f_2, f_3$ 
approximately fulfill $f_i = i\cdot f_1$. For each lattice the ground frequency $f_1$ can be obtained via the renormalization constant $Z(a)$, which in turn can be determined by examining the collective variable $Q_\text{cont}$. 

While instanton-updates and several other methods would also work here, Metadynamics is an ansatz which also seems promising for 4-dim. SU(3) theory \cite{Eichhorn:2021ccz, Eichhorn:2022wxn}. In the future, we plan to investigate bias potentials in that theory analogously to the way presented here. Additionally, in hopes of improving the effective sample size, we plan to experiment with different building strategies, such as well-tempered Metadynamics \cite{Barducci_2008} or other dynamical building methods. More generally, we intend to look into using other CVs in addition to $Q_\text{cont}$. What would also be interesting to see is if one can extract modes from the Markov chain which couple to the autocorrelation time by use of the generalized eigenvalue problem. Should one find observables with larger autocorrelation times than those of $Q$, one could (amongst other things) customize interesting CVs for Metadynamics.


\bibliographystyle{JHEP}
\setlength{\bibsep}{0pt plus 0.9ex}
\bibliography{literature.bib}

\end{document}